\documentclass[11pt]{article}
\usepackage{jheppub}
\usepackage{amsmath,amssymb,amsfonts,graphicx,slashed,color, mathtools, amsthm}
\usepackage{comment}
\usepackage{enumerate}
\usepackage{float}
\usepackage{caption}
\usepackage{subcaption}
\usepackage{pdflscape}

\newcommand{\be}{\begin{equation}}
\newcommand{\ee}{\end{equation}}
\newcommand{\bea}{\begin{eqnarray}}
\newcommand{\eea}{\end{eqnarray}}
\newcommand{\beas}{\begin{eqnarray*}}
\newcommand{\eeas}{\end{eqnarray*}}
\newcommand{\ba}{\begin{array}}
\newcommand{\ea}{\end{array}}

\title{Holographic motivations and observational evidence for decreasing dark energy}

\author[1]{Mark Van Raamsdonk,}
\author[2]{Chris Waddell}

\affiliation[1]{Department of Physics and Astronomy, University of British Columbia,\\
6224 Agricultural Road, Vancouver, B.C.\ V6T 1Z1, Canada}
\affiliation[2]{Perimeter Institute for Theoretical Physics, 31 Caroline St N, Waterloo, ON\ N2L 2Y5, Canada.}

\emailAdd{mav@phas.ubc.ca}
\emailAdd{cwaddell@perimeterinstitute.ca}

\abstract{Negative lambda gravitational effective field theories dual to holographic CFTs have potentially realistic cosmological solutions. Generic cosmological solutions of these effective field theories have scalar field evolution that can lead to a period of accelerated expansion when the scalar field is at positive values of its potential (Fig. 1). If such a model describes our universe, significant evolution of dark energy is expected over a Hubble time as the scalar descends from positive to negative values of its potential towards the AdS extremum. Our recent observational study 2305.04946  based on supernova and baryon acoustic oscillation (BAO) observations suggests that significant evolution of dark energy associated with a descending scalar field may be preferred by data (Fig. 2). Taking a linear approximation to the scalar potential around the present value, a standard likelihood analysis gives an $e^{- \chi^2/2}$ distribution in which $dV/dt$ is presently negative in $99.99 \%$ of the distribution, with a mean fractional variation of the potential of $36 \%$ over the period $z \lessapprox 2$ over which supernova data is available. In this note, we review these theoretical and observational results and provide an update on the question of how the physics of these cosmological solutions can be related to the physics of the underlying CFT.
}

\begin{document}

\maketitle

\section{Introduction}

A particularly exciting and compelling feature of the universe that we inhabit is that it exists. For a theoretical physicist, an evident corollary is that there also exists some complete, fully consistent mathematical model that provides a precise description of this universe. Finding such a model is one of the most important targets for quantum gravity research. 

A more modest target is to find and study fully consistent mathematical models that describe other possible cosmological spacetimes. Here, we mean universes that are approximately homogeneous and isotropic and expand from an initial big bang, but which might differ from our universe in details such as the spacetime dimension or the gravitational effective field theory.

The AdS/CFT correspondence (holography) provides us with a set of examples where we do have a complete mathematical model that encodes the quantum gravitational physics in a variety of interesting spacetimes. Thus, it is natural to ask whether we can use holography to describe cosmological spacetimes.

In the usual examples of holography, we have a CFT or more general quantum field theory and some $\Lambda < 0$ gravitational effective field theory associated with this. In the Lorentzian case, various states of the QFT describe various possible spacetimes that are solutions of the associated gravitational EFT with some fixed asymptotic behavior related to the UV behaviour of the QFT.\footnote{More complex CFT states can also describe spacetimes that include regions described by a different gravitational effective theory. See \cite{Simidzija:2020ukv} for example.} These spacetimes can have gravitating matter and radiation, black holes, etc., but they are always {\it asymptotically empty} and asymptotically AdS in the case when the holographic quantum system is a CFT. They are therefore quite different from cosmological spacetimes, for which we assume that the universe is filled uniformly with matter and/or radiation. 

On the other hand, precisely the same $\Lambda < 0$ gravitational effective field theories that have these asymptotically AdS solutions also have cosmological solutions. Because of the negative cosmological constant, these are typically big-bang / big-crunch cosmologies. Since the physics of the asymptotically AdS solutions is completely encoded in the physics of the corresponding CFT, it seems plausible that the physics of these cosmological solutions can also be related in some way to the physics of this CFT. One goal of this note is to summarize recent work \cite{Cooper:2018cmb, Antonini2019,VanRaamsdonk:2021qgv, Antonini:2022blk, Antonini:2022xzo} by us and collaborators attempting to elucidate how to extract cosmological physics from holographic CFTs. See \cite{Maldacena:2004rf, McInnes:2004nx,Usatyuk:2024mzs,Usatyuk:2024isz,Freivogel:2005qh,Banerjee:2018qey} for related work and \cite{Banks:2001px,Strominger:2001pn,Alishahiha:2004md,Gorbenko:2018oov,Coleman:2021nor,Freivogel2005,McFadden:2009fg,Banerjee:2018qey,Susskind:2021dfc} for various other approaches to holographically understand cosmological physics holographically.

\begin{figure}
    \centering
    \includegraphics[scale=0.5]{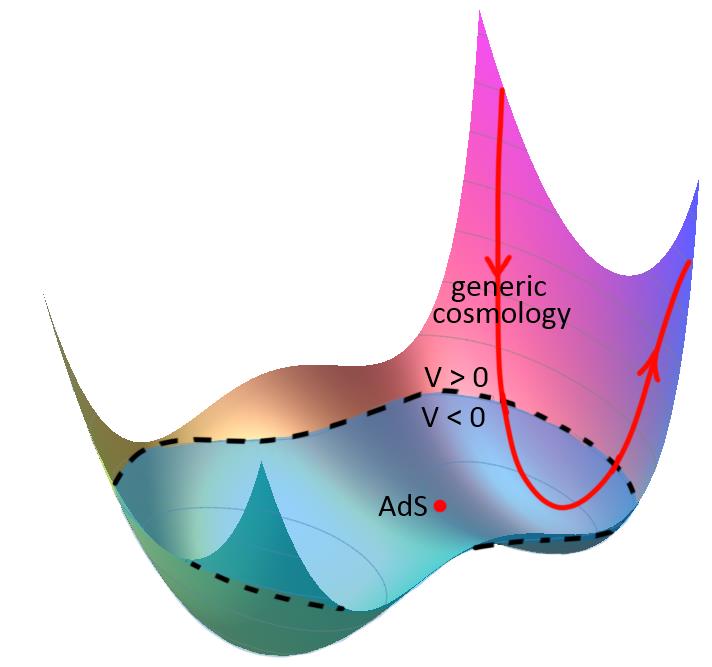}
    \caption{Schematic of the scalar potential for a typical holographic effective field theory. The red dot corresponds to an AdS extremum associated with a holographic CFT with relevant and irrelevant scalar operators. The red trajectory corresponds to a generic cosmological solution with an expanding phase that may include acceleration, followed by contraction to a big crunch. The scalar evolution is damped during expansion and anti-damped during contraction.}
    \label{fig:ScalarCosm}
\end{figure}

Studying cosmologies based on $\Lambda < 0$ effective field theories likely has a lot to teach us about the quantum gravity description of cosmology in general, just as studying AdS black holes has led to great insights into black hole physics and quantum gravity in general. However, a significant point that we would like to emphasize is that $\Lambda < 0$ cosmologies are not necessarily unrealistic. Indeed, as we review below, the $\Lambda < 0$ effective field theories associated to holographic CFTs typically have scalar fields with potentials that permit non-trivial scalar evolution on cosmological timescales \cite{Antonini2022,VanRaamsdonk:2022rts}. Cosmological solutions may generically have periods of time where the scalars are in a positive region of their potential, and this can lead to an accelerating phase. It is possible that the current accelerating expansion of our own universe can be explained in this way. Figure \ref{fig:ScalarCosm} shows the typical evolution of a scalar field in the potential of a holographic EFT for a generic cosmological solution with matter and radiation.

If one of these holographic $\Lambda < 0$ EFTs does describe our universe, the most significant prediction for observations is that we should find the dark energy to be changing with time, likely decreasing as the scalar field rolls down its potential. This is an exciting prospect: we have a class of theoretical models that are UV-completable, can potentially explain the observed late time acceleration, and have a generic testable prediction, that dark energy should be changing (and likely decreasing) with time because of scalar evolution. 

\begin{figure}
    \centering
    \includegraphics[scale=0.5]{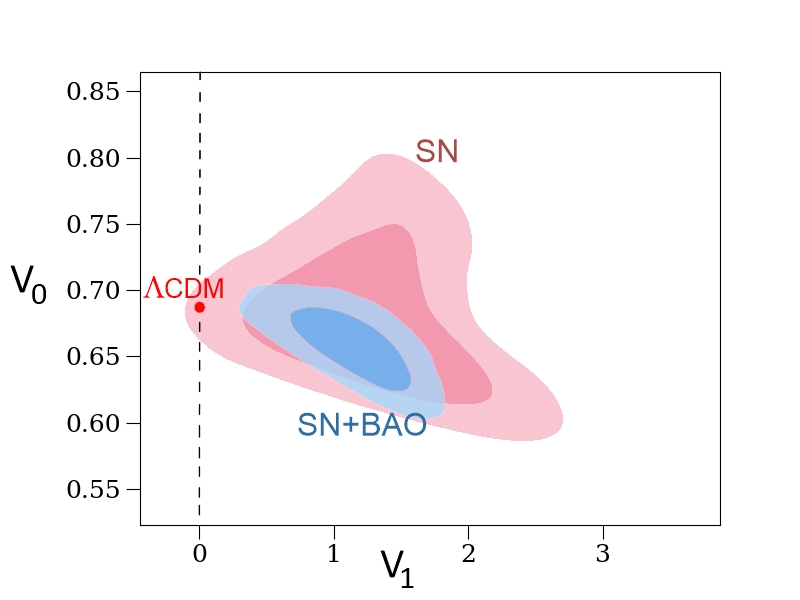}
    \caption{Distribution of parameter values $V_0$ and $V_1$ in a linear approximation $V(\phi) = V_0 + V_1 \phi$ to the scalar potential in the convention that $\phi=0, \dot{\phi} \le 0$ presently. The $\exp(-\chi^2/2)$ likelihood function is computed by comparing model predictions to redshift vs brightness data in the Pantheon+ supernova catalogue (68\% and 95\% contours shown in pink) and to combined supernova and BAO data (blue). Best fit $\Lambda$CDM model is indicated in red. Positive $V_1$ (corresponding to decreasing potential energy at present in our conventions) accounts for $\approx 99\%$ of the distribution with only supernova data and $\approx 99.99 \%$ with supernova and BAO data. Here, $V$ is normalized so that $V_0 = \Omega_\Lambda$ in a $\Lambda$CDM model with $V_1 = 0$, and our conventions for $V_1$ (setting $8 \pi G/3 = 1$) are such that an order one value typically corresponds to an order one decrease in $V$ over a Hubble time. ({\it From \cite{VanRaamsdonk:2023ion}.})}
    \label{fig:V0V1}
\end{figure}

As we review below, the generic expectation from this class of models is that the scalar potential should change by an order one fraction in a Hubble time. The resulting significant evolution of dark energy should almost certainly be observable with near future observations. Indeed, it is natural to ask whether such a significant evolution of dark energy might already be ruled out by available data. In order to investigate this, we decided to perform an independent analysis constraining the parameters of a scalar potential model using the latest available data for two different probes of the recent scale factor evolution, type Ia supernova redshift vs brightness and the imprint of early universe baryon acoustic oscillations on large scale structure \cite{VanRaamsdonk:2023ion}.\footnote{Typical cosmological analyses investigate the prospect of changing dark energy in the context of specific modifications of the Friedmann equation that model the $w$ parameter of dark energy as a constant or a linear function of $a$. These models don't necessarily arise from some consistent effective field theory, and they don't provide a good approximation to actual scalar potential models, for which $w(a)$ can be significantly non-linear when there is a significant change in dark energy.}  

While we don't have a specific prediction for a global potential from holography, the data we utilize explores the relatively recent cosmological past ($z < 2$). During this time, the scalar would have explored only a limited region of its potential, so we choose to work with a linear approximation to the potential around its present value. This approximation is certainly valid for some range of time around the present day, but not necessarily over the whole range of time for which supernova and BAO data are available. On the other hand, if there is significant evolution of dark energy, one might expect that such a linear potential model would provide a significantly better fit than $\Lambda$CDM even if it is not precisely the right potential.

The results were surprising to us. As shown in Figure \ref{fig:V0V1}, in the context of these linear potential models, the best fit $\Lambda$CDM model lies well outside the 95 $\%$ confidence interval associated with the $e^{- \chi^2/2}$ likelihood distribution.\footnote{We review the definition of $\chi^2$ below.} Approximately $99.99 \%$ of the distribution has $dV/dt < 0$ (decreasing scalar potential at the present time), with the best fit models having an order 1 fractional change in $V$ during the period of evolution over which supernova data are available. The results should be treated with some caution: the distribution is non-Gaussian, and the difference in $\chi^2$ between the best fit $\Lambda$CDM model and the best fit model overall ($\Delta \chi^2 \approx 3.7$) is not as large as would be expected given how far out $\Lambda$CDM sits on the tail of the distribution. It should almost certainly become clear with improved data in the near future whether the suggested variation in dark energy is real. The recent DESI analysis \cite{DESI:2024mwx} using newer supernova and BAO data also finds some preference for time-varying dark energy, though the $w_{0}$-$w_{a}$ model they consider has $w < -1$ for a significant amount of the evolution and may be unphysical. We hope to repeat our analysis with these newer data sets when they become available.

The most important takeaway here is that there seems to be plenty of room for cosmological models based on gravitational $\Lambda < 0$ effective field theories. The observational results do not specifically provide  evidence for holographic effective theories -- these are just specific examples of quintessence models \cite{Peebles:1987ek,Ratra:1987rm,Caldwell:1997ii,Dutta:2018vmq,Visinelli:2019qqu,sen2021cosmological} where the potential goes negative. However, it is holography that provides the motivation for such theories being natural to consider, and in particular for the naturalness of scalar field variation on cosmological timescales.  

\subsubsection*{Outline}

Here is the plan for the remainder of the paper. In Section \ref{sec:holo_EFTs}, we review some basics of the $\Lambda < 0$ gravitational effective theories associated with holographic CFTs and their cosmological solutions. We review the arguments that these generically have scalar evolution on cosmological timescales. In Section \ref{sec:obs}, we summarize the observational results from \cite{VanRaamsdonk:2023ion} and also comment on the relation to the recent DESI results that have also been taken to suggest time-dependent dark energy. In Section \ref{sec:cft}, we discuss the question of how the physics of these cosmological solutions may be related to the physics of associated CFTs. In Section \ref{sec:dis}, we briefly describe some open questions. 

\section{Cosmology from holographic effective field theories} \label{sec:holo_EFTs}

For a given holographic CFT, the associated  gravitational effective field theory describing the bulk physics can be deduced from basic properties of the CFT. The light fields in the gravitational theory are in one-to-one correspondence with the low-dimension operators in the CFT. For example, the stress-energy tensor corresponds to the metric field, while any current operators correspond to gauge fields.

For each scalar primary operator ${\cal O}_i$ in the CFT of dimension $\Delta_i$, we have a scalar field $\phi_i$ in the associated gravitational EFT with mass given by $m_i^2 \ell_{AdS}^2 = \Delta_i (\Delta_i -3)$. There will be some potential $V(\phi_i)$ for these scalars (and possibly a non-trivial metric on the scalar target space) with a negative extremum at a point that we can choose to be $\phi_i=0$. The value of the potential here is the negative cosmological constant $\Lambda$, whose value is related to the number of degrees of freedom in the CFT. Scalars associated with relevant operators correspond to directions away from $\phi_i=0$ with negative second derivative for V, while those associated with irrelevant operators correspond to directions with a positive second derivative. Globally, the potential may be complicated due to scalar interaction terms. We expect that the full potential has regions with both negative and positive values and is overall bounded below for stability. Figure \ref{fig:ScalarCosm} shows a schematic of such a potential with two scalar fields.

Cosmological solutions have a metric that (approximately) takes the FRW form
\begin{equation} \label{eq:FRW}
    ds^2 = -dt^2 + a^2(t) d \Sigma^2 \; ,
\end{equation} 
where the spatial metric $d \Sigma^2$ can be flat, positively curved, or negatively curved.
The symmetries preserved by this ansatz also admit time-dependent scalar fields
\begin{equation} \label{eq:homogeneous_scalar}
    \phi_i = \phi_i(t) \; .
\end{equation}
Generally, we also have approximately homogeneous and isotropic matter and radiation.

In a typical cosmological solution for a $\Lambda < 0$ EFT, matter and radiation dilute as the universe expands after the big bang and the scalar potential energy or cosmological constant becomes increasingly important. There may be a temporary period of acceleration if there is a scalar descending from positive values of its potential, but ultimately negative potential values are reached and the universe recollapses. At the turnaround point $\dot{a} = 0$, the Friedmann equation implies that the total energy density vanishes, so the scalar potential energy has precisely the same magnitude as the remaining sources of energy. An interesting feature of these cosmologies is that unlike the $\Lambda > 0$ case, the potential energy is of a similar order of magnitude to the remaining sources of energy for most of the history of the universe, so there is no cosmological coincidence problem.

\subsection{Naturalness of scalar evolution in cosmology with holographic EFTs}

An important point about the scalar potentials in holographic EFTs is that the scale of the variation of the potential (i.e. the scale of the derivatives) is such that the natural timescale for the variation of the scalars is the cosmological timescale set by the AdS length. 

In typical examples of AdS/CFT, we can write the action of gravity and the scalars as 
\be
\label{scalaraction}
S = \frac{1}{2\kappa} \int d^4x \sqrt{g}(R -  g^{ab}\partial_a \phi \partial_b \phi - \frac{6}{\ell_{AdS}^2} V(\phi)) 
\ee
where $\kappa = 8 \pi G$. In this expression, the scalar field and its potential $V(\phi)$ are dimensionless and the parameters in the potential 
\be
\label{eq:Vpot}
V(\phi) =   -1 + \frac{1}{2} \hat{m}^2 \phi^2 + V_{int}(\phi) \; .
\ee
are typically of order 1.\footnote{For the mass parameter, this follows from the usual relation  $\hat{m}^2 = m^2 \ell_{AdS}^2 = \Delta(\Delta - 3)$ between scalar operator dimensions and scalar field masses. For the interactions, this is related to usual large $N$ scaling of CFT correlation functions.} In this case, $\ell_{AdS}$ is the only scale appearing in the equations of motion so this sets the timescales for variation of both the metric (i.e. the Hubble scale in the cosmology) and the scalars.

While there are cosmological solutions of the effective theory where the scalars just sit at an extremum of their potential, generic solutions have scalar evolution, and this evolution typically will have an order 1 fractional variation of the potential over cosmological timescales.\footnote{During the expanding phase of the universe, the evolution of the scalar field is governed by the equation for a particle moving in a potential equal to the scalar potential with damping coefficient $3 H$ where $H = \dot{a}/a$ is the Hubble parameter. }  A natural class of solutions has the scalar descending from positive values in the early universe to negative values at late times; these solutions can have an accelerating phase and may be relevant for our observed cosmology. For cosmologies arising from EFTs with a CFT dual,  a dynamical scalar field that descends from positive to negative values of its potential is the only way to explain acceleration; thus, time-dependent dark energy can be seen as a generic prediction of these models.

Rescaling to particle physics conventions, the scalar action becomes
\be
\small{S_\phi = \int d^4 x \sqrt{g} \left[-\frac{1}{2} g^{ab}\partial_a \phi_p \partial_b \phi_p - \frac{\hat{m}^2}{\ell_{AdS}^2}  \phi_p^2 -\frac{1}{\kappa \ell_{AdS}^2} V_{int}( \sqrt{\kappa} \phi_p )\right]} \; .
\ee
Here the scalar mass $m \sim 1/\ell_{AdS}$ and coupling $g \sim \ell_P/\ell_{AdS}$ might both seem unnaturally small from a particle physics point of view, but the key point is that there is already a large hierarchy of scales $\ell_{AdS}/\ell_P \gg 1$ related to the large number of degrees of freedom in the underlying CFT, and the masses and couplings are suppressed by this same large number.

A more detailed discussion on these naturalness issues and the genericity of scalar evolution in cosmology is presented in \cite{VanRaamsdonk:2022rts}. In the next section, we discuss observational evidence for this class of cosmological models, returning to the question of how such models might arise microscopically from CFT physics in Section \ref{sec:cft}.

\section{Observational evidence for scalar evolution} \label{sec:obs}

In this section, we review the results of \cite{VanRaamsdonk:2023ion} where we use observational data to test the core premise of the above theoretical considerations: that our universe could plausibly be described by a $\Lambda < 0$ gravitational EFT, with fields evolving from a presently positive value of their potential. 

In such effective field theories, spatially flat cosmological solutions of the form in (\ref{eq:FRW}), (\ref{eq:homogeneous_scalar}) are described by the Friedmann and scalar equations\footnote{We are assuming a flat field space metric. }
\be \label{eq:eoms}
H^2 = {8 \pi G \over 3}\left[\rho + {1 \over 2} \dot{\vec{\phi}}^2 + V(\vec{\phi}) \right] \: , \qquad
\ddot{\phi}_{i} + 3 H \dot{\phi}_{i} +  \partial_{i} V(\vec{\phi}) = 0 \: ,
\ee
where an overdot denotes a $t$ derivative and $H = \dot a/a$ is the Hubble parameter, and where $\rho$ is the energy density excluding that associated with the scalar field. 
Given that the contribution to the total energy density from radiation is negligible for the recent expansion history, we typically consider $\rho$ to consist only of pressureless matter. Henceforth working in conventions where the current scale factor is $a_{0} = 1$, we may write
\begin{equation}
    \rho = 
    {3 H_0^2 \Omega_M \over 8 \pi G} {1 \over a^3} \: ,
\end{equation}
where 
$H_{0}$ is the current value of the Hubble parameter, and $\Omega_{M}$ is the current fraction of the total energy density in matter.

For comparison with observational data, one would like to relate the scale factor evolution to cosmological redshifts and distances. The redshift $z = \frac{1}{a} - 1$ is precisely the multiplicative factor describing the optical redshift of a photon emitted at scale factor $a$. The comoving distance between a source which emits a photon at redshift $z$ (time $t(z)$) and an observer which receives the photon at present is 
\begin{equation}
    d_{C}(z) = \int_{t(z)}^{0} \frac{dt'}{a(t')} = \int_{0}^{z} \frac{dz'}{H(z')} \: ,
\end{equation}
providing an integrated probe of the time-dependence of $a(t)$. 
In a flat cosmology, $d_{C}(z)$ is the same as the transverse comoving distance $d_{M}(z)$, defined as the ratio of the comoving distance between two objects at redshift $z$ and their apparent (small) angular separation.

\subsection{Observational data}

The coupled equations (\ref{eq:eoms}) determine the scale factor evolution given $\Omega_{M}$ and $V(\phi)$, and initial conditions; conversely, scale factor observations can constrain these inputs.
We consider two direct observational probes of the recent scale factor evolution: type Ia supernovae (SNe Ia), and imprints of baryon acoustic oscillations (BAOs) in large scale structure. 


Type Ia supernovae, thermonuclear explosions of stars in binary systems which accrete sufficiently large mass, provided the first evidence that the universe is undergoing an accelerated expansion \cite{SupernovaSearchTeam:1998fmf, SupernovaCosmologyProject:1998vns}, and remain a gold standard probe for constraining the dynamics of dark energy. They are ``standard candles", thought to have a relatively standardized peak intrinsic brightness, with any variability empirically correctable using detailed information about the time-dependence of this brightness. 
With knowledge of the absolute brightness, often expressed in dimensionless logarithmic units as an absolute magnitude $M$, the apparent magnitude $m$ of a SN Ia can thus be related to a distance measure;\footnote{Calibrating the distance-brightness relationship, i.e. determining $M$, requires additional input from other distance indicators; in our analysis, this comes from Cepheid variable stars in galaxies hosting SNe Ia.} whereas in flat spacetime the brightness would be expected to decrease as the inverse square of the distance from a source, in a cosmological spacetime this distance should be replaced by a scale factor-dependent quantity known as the luminosity distance
\begin{equation}
    d_{L}(z) = (1+z) d_{M}(z) \: , \qquad m - M = 5 \log_{10} \left( \frac{d_{L}}{10 \: \text{pc}} \right) \: .
\end{equation}
Combining redshift and brightness information then provides a direct measure of the recent expansion. 
The catalogue of SNe Ia observations that we use is the Pantheon+ sample \cite{Pan-STARRS1:2017jku, Scolnic:2021amr},\footnote{Recently, a large number of new high-redshift SNe Ia observations has been collected by the Dark Energy Survey in \cite{DES:2024tys}. When combined with BAOs, these data yielded the highest significance in the favourability of a model with evolving dark energy over $\Lambda$CDM in the DESI analysis \cite{DESI:2024mwx}. } 
consisting of 1701 spectroscopically confirmed SNe Ia in the approximate range $0 < z \lesssim 2.3$.
The Pantheon+ data has been used to constrain cosmological parameters in $\Lambda$CDM and simple extensions thereof \cite{brout2022pantheon+}. 

Baryon acoustic oscillations are another crucial dark energy probe, which for example underlie the recently announced cosmological constraints from the DESI collaboration \cite{DESI:2024mwx}. BAOs provide a ``standard ruler"; 
the comoving scale at which correlation functions of inhomogeneities are expected to be peaked is fixed by the comoving sound horizon at the time that a given species decoupled from thermal equilibrium,
and this feature is visible in both the cosmic microwave background (CMB) and the distribution of galaxies.\footnote{In addition to galaxies, some large scale structure measurements involve the distribution of quasars, as well as Lyman-$\alpha$ forest auto-correlation and cross-correlation with quasars. } 
The fact that we can track this feature as a function of redshift in the case of galaxies allows us to put constraints on the scale factor evolution.
Concretely, assuming some fiducial value of the comoving sound horizon at decoupling $r_{d}$, observations of the BAO feature at redshift $z$ along the line of sight fix the Hubble scale via $\delta z_{\text{BAO}} = H(z) r_{d}$, while observations of the transverse scale fix the comoving distance $\delta \theta_{\text{BAO}} = \frac{r_{d}}{d_{M}(z)}$.
For our analysis, we follow \cite{brout2022pantheon+} in drawing on BAO data from a number of surveys \cite{Ross:2014qpa, BOSS:2016wmc, eBOSS:2020lta, eBOSS:2020hur, eBOSS:2020fvk, eBOSS:2020uxp, eBOSS:2020gbb, eBOSS:2020tmo};\footnote{New data has recently been made available in \cite{DESI:2024mwx}. } individual surveys each comprise independent sets of hundreds of thousands of galaxies, though the result of the analyses \cite{Ross:2014qpa, BOSS:2016wmc, eBOSS:2020lta, eBOSS:2020hur, eBOSS:2020fvk, eBOSS:2020uxp, eBOSS:2020gbb, eBOSS:2020tmo} for a given survey, which we take as input, is typically the value of one or two distance measures at a single effective redshift.  

\subsection{Methodology}

Given a cosmological model and a data set, we would like to compare the two in order to make inferences about the model parameters. To do so, we  define a likelihood function
\begin{equation}
    \mathcal{L}(\mathcal{D}) = P(\mathcal{D} | \theta) \: ,
\end{equation}
namely a probability distribution for data set $\mathcal{D}$ given some fixed model parameters $\theta$. In the case of the SN Ia data discussed above, we follow the Pantheon+ analysis of cosmological constraints \cite{brout2022pantheon+} and define a likelihood function based on the $\chi^{2}$ statistic $\mathcal{L} \propto e^{- \chi^{2} / 2}$, where\footnote{A subtlety is that, for observations of SNe Ia in galaxies hosting a Cepheid variable being used to calibrate the distance-brightness relation, the contribution to $\chi^{2}$ is modified; see \cite{VanRaamsdonk:2023ion} for details.}
\begin{equation}
    \chi^{2} = (\mu_{\text{data}} - \mu_{\text{model}})_i C^{-1}_{ij} (\mu_{\text{data}} - \mu_{\text{model}})_j \: . 
\end{equation}
Here, the quantity $\mu_{\text{data}, i} = m_{\text{data}, i} - M$ comes from the $i^{\text{th}}$ brightness measurement, $\mu_{\text{model}, i} = 5 \log_{10} (d_{L}(z_{i}) / (10 \: \text{pc}) )$ comes from the $i^{\text{th}}$ redshift measurement and the model, and $C_{ij}$ is an appropriate covariance matrix capturing the statistical and systematic errors in the data. 
The likelihood functions associated to BAO data are based on those provided by \cite{Ross:2014qpa, BOSS:2016wmc, eBOSS:2020lta, eBOSS:2020hur, eBOSS:2020fvk, eBOSS:2020uxp, eBOSS:2020gbb, eBOSS:2020tmo}.\footnote{An implementation of these likelihoods for $\Lambda$CDM can be found in the CosmoSIS framework \cite{Zuntz:2014csq}. } 

Given a prior distribution $P(\theta)$, Bayes' Theorem may be invoked to determine the posterior distribution, which allows $\theta$ to be estimated given $\mathcal{D}$,
\begin{equation} \nonumber
    P(\theta | \mathcal{D}) = \frac{P(\mathcal{D} | \theta) P(\theta)}{P(\mathcal{D})} \propto P(\mathcal{D} | \theta) P(\theta) \: .
\end{equation}
The best fit parameters may be computed via $\textnormal{argmax}_{\theta} P(\theta | \mathcal{D})$. One may also marginalize over some subset of parameters to obtain a probability distribution with respect to the remaining parameters; marginalizing over all but one or two parameters yields the parameter values and their uncertainties in Table \ref{tab:params_BAO}, and the contour plots in Figure \ref{fig:LinearPotential_Results_SNBAO}. 

In practice, given the computational expense involved in computing a likelihood $\mathcal{L}$, one requires an efficient method for sampling from the posterior probability distribution. This is provided by Markov Chain Monte Carlo (MCMC) methods, the prototypical example being the Metropolis-Hastings algorithm.
Typically, an MCMC algorithm proceeds by choosing an initial set of $n$ points for the sample, generating a new set of $n$ points based on a fixed protocol (e.g. a random walk), and either retaining or discarding the new points based on some criterion 
designed to favour points with higher likelihood. 
The distribution of points in the sample generated this way converges to the posterior probability distribution in the limit of large statistics. 

For additional information and context regarding our methodology, and the results we turn to next, please consult \cite{VanRaamsdonk:2023ion}. 

\subsection{Results: Linear potential} \label{sec:lin}

The full potential associated with any particular holographic effective field theory is likely complicated. However, it is plausible that over the recent cosmological past, the scalar fields have only been exploring a small part of the potential. We can approximate the scalar potential near the present value via a Taylor approximation, and for our analysis we keep only the constant and linear terms in this approximation. We also focus on the case of a single scalar; this can be taken to model the present direction of evolution in the scalar target space.\footnote{The reduction to a single scalar is most appropriate if the scalar is moving along a gradient, but this is made plausible by the Hubble damping in the scalar equation.}

We may now apply the analysis described above to this model with a single minimally coupled scalar field subject to a linear potential. The scalar action is\footnote{Since a constant rescaling of time in the equations of motion (\ref{eq:eoms}) is equivalent to an overall rescaling of the potential, we can think of $H_{0}$ as setting the scale of the potential, so we explicitly factor this out when defining parameters $V_{0}, V_{1}$. }
\begin{equation}
    S_{\phi} = - \int d^{4} x \: \sqrt{|g|} \left( \frac{1}{2} \partial_{\mu} \phi \partial^{\mu} \phi + V(\phi) \right) \: , \qquad \frac{V(\phi)}{H_{0}^{2}} = V_{0} + V_{1} \phi \: ,
\end{equation}
where we are working in units $\frac{8 \pi G}{3} = 1$ (meaning that $V(\phi) / H_{0}^{2}$ varies by an $O(1)$ amount when $\phi$ varies by a Planckian amount when $V_{1} = O(1)$). 
Without loss of generality, we set the conventions that the current value of $\phi$ is zero, and that $\phi$ is currently moving toward negative values, so that $V_{1} > 0$ corresponds to a currently decreasing dark energy. 

In addition to the cosmological parameters $(H_{0}, \Omega_{M}, V_{0}, V_{1})$, we also take the absolute magnitude of type Ia supernovae $M$ and the comoving sound horizon $r_{d}$ to be free parameters in the MCMC. We assume uniform priors, though with the positive energy condition $\Omega_{M} > 0$ and the constraint $\Omega_{M} + V_{0} < 1$, equivalent to the condition that the scalar kinetic energy is positive by equation (\ref{eq:eoms}). 

\begin{figure}
    \centering
    \includegraphics[width=15cm]{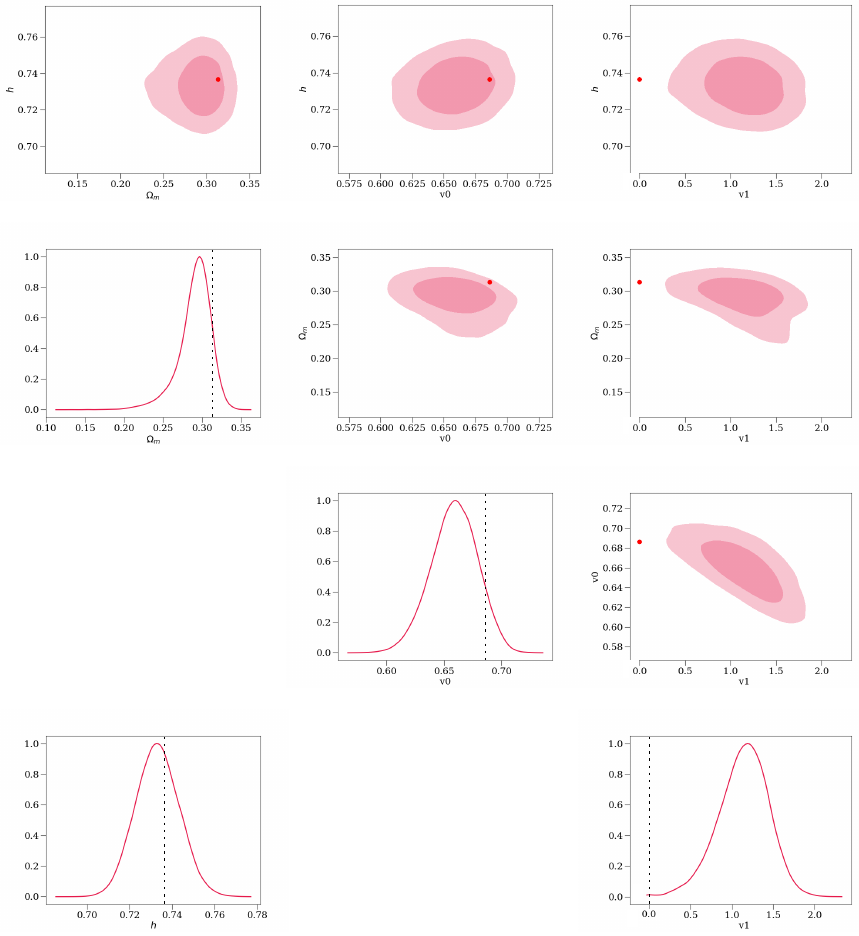}
    \caption{ Distribution of parameter values $h$, $\Omega_M$, $V_0$, and $V_1$ for a linear potential scalar field model, using Pantheon+ data calibrated with SH0ES Cepheid distances and BAO data (68\% and 95\% contours shown). We define $h$ by $H_{0} = 100 \: h \: \text{km} \: \text{s}^{-1} \: \text{Mpc}^{-1}$. The $\Lambda$CDM model corresponds to $V_0 = 1 - \Omega_M$, $V_1 = 0$; the best fit $\Lambda$CDM parameter values are demarcated by a red dot or a dashed black line in the figure. (\textit{From \cite{VanRaamsdonk:2023ion}.})}
    \label{fig:LinearPotential_Results_SNBAO}
\end{figure}

In Figure \ref{fig:LinearPotential_Results_SNBAO}, we show the marginalized distributions for the cosmological parameters. Perhaps the most interesting feature of these results is that approximately $99.99\%$ of the entire distribution of sampled values is supported on $V_{1} > 0$, implying that $\Lambda$CDM is significantly disfavoured (beyond the 3.5$\sigma$ level) within this class of models.
We find the marginalized average value $V_{1} \approx 1.1$; relatedly, we find that the typical ratio between the range of scalar evolution and the current value of the scalar potential is an $O(1)$ quantity, $\langle \text{Range}[V(\phi)]/V_{0} \rangle \approx 0.36$, and the average time into the future at which the scalar field will descend to negative values of the potential is approximately $2.1$ Hubble times. 
The marginalized average values of the remaining cosmological parameters, recorded in Table \ref{tab:params_BAO}, are broadly in accord with analyses based on the flat $\Lambda$CDM model, e.g. \cite{brout2022pantheon+}.

Although the posterior distribution appears to have negligible support on $V_1 < 0$, we find that the difference between best fit $\chi^2$ values (more precisely, $-2 \ln \mathcal{L}$ values) in the $\Lambda$CDM model and the linear potential model is relatively small,
\begin{equation}
    \Delta \chi^{2} = \chi^{2}_{\text{best fit, } \Lambda\text{CDM}} - \chi^{2}_{\text{best fit, linear pot.}} \approx 3.7 \: .
\end{equation}
For a Gaussian distribution, $\Delta \chi^2$ would be much larger for points so far out on the tail of the distribution, suggesting that the distribution here is highly non-Gaussian, with a very localized peak surrounding the best fit $\Lambda$CDM model and a much broader peak, accounting for most of the distribution, around the best fit linear potential model.\footnote{This is perhaps partially due to the fact that, when $V_{1}$ is sufficiently small, $\Omega_M$ and $V_0$ appear to be strongly correlated for high likelihood models, suggesting a reduction of the effective dimension of the parameter space.} Thus, while the standard likelihood analysis favours linear potential models with significant scalar evolution, alternative model selection criteria which do not account for parameter space volumes do not demonstrate such a preference. The additional data required to definitively distinguish between these alternatives should be available in the coming years, and in fact may appear in very recent or pending data releases \cite{DES:2024tys, DESI:2024mwx}. 

\begin{table}
\begin{center}
\begin{tabular}{|| c | c | c | c ||}
 \hline
 $h$ & $\Omega_{M}$ & $V_{0}$ & $V_{1}$ \\ [0.5ex] 
 \hline\hline
 $0.733 \pm 0.010$ & $0.291^{+0.023}_{-0.013}$ & $0.659^{+0.021}_{-0.019}$ & $1.1 \pm 0.3$ \\
 \hline
\end{tabular}
\caption{Summary of marginalized parameter constraints for the linear potential model, based on supernova and BAO data. Here $h$ is defined by $H_{0} = 100 \: h \: \text{km} \: \text{s}^{-1} \: \text{Mpc}^{-1}$. More detailed information about the distribution of parameter values can be found in Figure \ref{fig:LinearPotential_Results_SNBAO}. }
\label{tab:params_BAO}
\end{center} 
\end{table}

An important distinction between our results and those of the recent DESI analysis \cite{DESI:2024mwx} that have also been interpreted to favour dynamical dark energy is that we are working directly with a consistent effective field theory, while the $w_{0}$-$w_{a}$ model studied in \cite{DESI:2024mwx} is a particular modification of the Hubble equation that takes the equation of state parameter $w = p/\rho$ for dark energy to be a linear function of the scale factor. It's not likely that this can naturally be realized by some effective field theory, and in fact the preferred models of \cite{DESI:2024mwx} have $w < -1$ for a significant part of their evolution. This may be associated with various instabilities and violations of energy conditions. 

\section{Connecting to the CFT} \label{sec:cft}

In this section, we discuss the question of how the physics of cosmological solutions of $\Lambda < 0$ gravitational theories associated with holographic CFTs might be related to the physics of those CFTs.

An apparent obstacle to relating the cosmological physics to the associated CFT is that these cosmological solutions do not have any asymptotically AdS region. Thus, the cosmology is not simply dual to a state of the CFT on some spacetime. In the following subsections, we describe two explicit ways that we can connect the cosmological physics to the CFT. The first is a ``local'' approach (similar to describing the static patch of de Sitter space) in which we aim to encode the physics of regions that contain the causal patch accessible to an observer but not necessarily the global spacetime. In the second approach, we aim to encode the full cosmological spacetime and also to describe a preferred state, similar to the Hartle-Hawking state.

\subsection{The local approach}

For $\Lambda < 0$ big-bang / big-crunch cosmologies we have both past and future singularities, with a finite proper time between them along any timelike trajectory. The causal diamond (i.e. the intersection between the causal past and the causal future) associated with any such trajectory has a finite spatial size; it is included in the domain of dependence of a finite spatial region. If we are interested in encoding the physics accessible to any single observer in the cosmology, it is only necessary to produce a holographic description of such a region. This is analogous to efforts to holographically describe the static patch of de Sitter space.

\begin{figure}
    \centering
    \includegraphics[width=10cm]{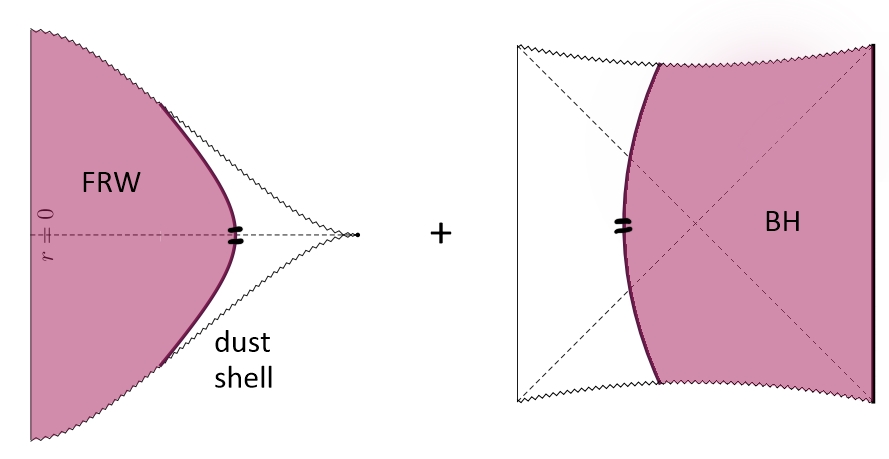}
    \caption{Causal diagram for a finite bubble of big-bang / big-crunch cosmology (left) serving as the interior region for a large AdS/Schwarzchild black hole. For general matter and radiation in the FRW region, the domain wall separating the two parts of the spacetime can be taken to be a dust shell.}
    \label{fig:FRW_BH}
\end{figure}

The description of such finite bubbles of $\Lambda < 0$ cosmology via a standard holographic setup was considered in \cite{Sahu:2023fbx}, following \cite{Freivogel2005} who considered $\Lambda > 0$ bubbles.
As described in detail in that paper, it is possible to embed such a finite ``bubble'' of cosmology into an asymptotically AdS spacetime by considering a solution with an empty AdS-Schwazschild exterior patched on to the FRW bubble interior along a domain wall that can be taken to be a thin shell of dust, as shown in Figure \ref{fig:FRW_BH}.

From the exterior perspective, this solution is a large AdS black hole. In most cases, the entropy of the resulting black hole is larger than the entropy of the matter and radiation in the cosmological region. This suggests that there can be a high-energy dual CFT state that faithfully encodes the cosmological physics. In \cite{Sahu:2023fbx}, we describe a possible Euclidean path integral construction of such a state.

For these cosmological bubble solutions, most or all of the cosmology region is behind the horizon of the black hole from the exterior perspective. Thus, understanding how the cosmological physics is encoded in the CFT is related to the challenging problem of understanding how behind-the-horizon physics is encoded. On the other hand, when we have a Euclidean construction of the state, it is much easier to extract the physics via this Euclidean picture. 

In this construction, the big bang is part of the past singularity of the black hole. So understanding the physics near the big bang singularity is a special case of understanding physics of black hole singularities using AdS/CFT. The cosmology perspective suggests that an interesting question (either for the cosmological bubbles or for simple two-sided black holes) is to understand better the state of the quantum fields near the past singularity, as this provides the initial state for the cosmology. It would be interesting to understand what sort of perturbations are naturally present, whether there is an arrow of time that emerges, etc... . 

The bubble construction has the ugly feature that homogeneity and isotropy are present only locally in the solutions. While it's impossible to know whether there is some empty asymptotically AdS region past the cosmological horizon in our own universe, this possibility seems far-fetched and unappealing! However, the important point is that the details of the exterior region of the spacetime should have no bearing on what happens in the cosmological region. We can think of this construction as one of many possible descriptions of the physics accessible to an observer in a fully homogeneous and isotropic cosmology.
This point was highlighted recently in \cite{VanRaamsdonk:2018zws, VanRaamsdonk:2020ydg,Simidzija:2020ukv}, where we showed that bubbles of an asymptotically AdS spacetime encoded by a holographic CFT can also be faithfully encoded in various other holographic CFTs or even in collections of disconnected BCFTs.\footnote{A related perspective is that the physics of the causal patch associated with a worldline in the cosmology has some associated algebra of observables \cite{Leutheusser:2022bgi}. In the present construction, this likely emerges as a subalgebra from the full algebra of observables of the CFT in the large $N$ limit. But the same algebra can also be described in other ways.}

\subsection{The global approach}

While a global description of the cosmological spacetime may not be strictly necessary, it would be unsatisfying if such a description did not exist. In this section, we discuss a possible route to understanding such a global description, following \cite{Maldacena:2004rf,Cooper2018,VanRaamsdonk:2020tlr,VanRaamsdonk:2021qgv,Antonini2022}.

\begin{figure}
    \centering
    \includegraphics[width=7cm]{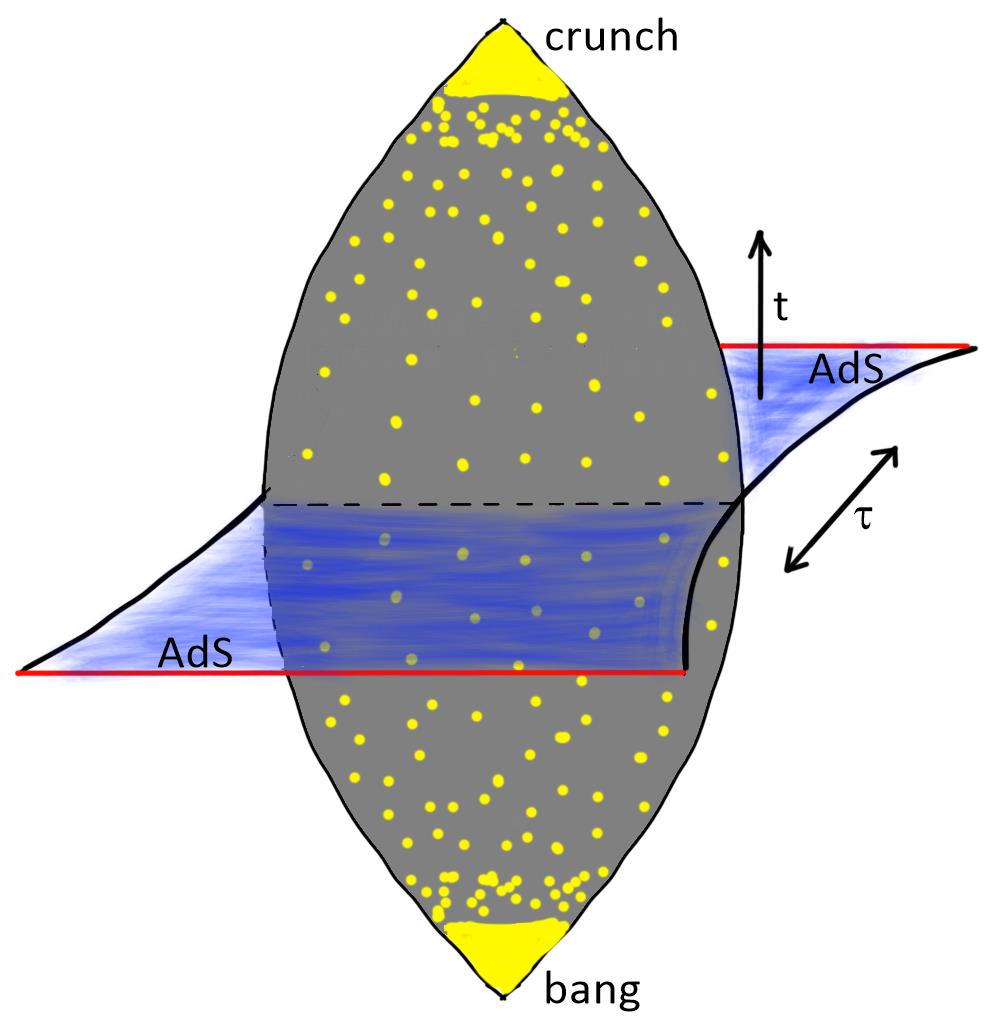}
    \caption{A time-symmetric big-bang / big-crunch cosmology and its Euclidean continuation to an asymptotically AdS Euclidean wormhole.}
    \label{fig:Lor-Euc}
\end{figure}

The starting point is the observation that many cosmological solutions of $\Lambda < 0$ gravitational effective field theories are time-symmetric at the level of background cosmology and that these time-symmetric geometries are asymptotically AdS for Euclidean time $\tau \to \pm \infty$, as depicted in Figure \ref{fig:Lor-Euc}. 

For example, the background solution for flat $\Lambda < 0$ radiation cosmology, with metric
\begin{equation}
    ds^2 = -dt^2 + \cos(2t/\ell_{AdS}) d \vec{x}^2 \: ,
\end{equation}
continues to the Euclidean solution
\begin{equation}
    ds^2 = d\tau^2 + \cosh(2t/\ell_{AdS}) d \vec{x}^2
\end{equation}
that can be described as a planar asymptotically AdS wormhole. 

The pair of asymptotically AdS regions for $\tau \to \pm \infty$ suggests the existence of a holographic description of the Euclidean spacetime involving a pair of three-dimensional Euclidean CFTs. Such a description could be used to define and compute observables in the Euclidean spacetime; observables in the cosmology could be obtained from these via analytic continuation. 

Such a Euclidean construction chooses a particular preferred state for the cosmology. We can think of the Euclidean construction as defining a gravitational path integral that prepares a state for the Lorentzian cosmology. This is similar to the Hartle-Hawking no-boundary proposal, but in our case, we have asymptotically AdS boundaries in the Euclidean past and future.

A key question is how to holographically obtain a wormhole solution starting with two Euclidean CFTs. If the wormhole provides a dominant saddle, the connectedness of the solution implies correlations between the two CFTs. The source of these correlations could either be an interaction or some type of ensemble.\footnote{Another way to think about the wormhole is that it may correspond to some contribution to the factorized partition function of the pair of CFTs that is insensitive to the precise microscopic details of these CFTs. Adding these other contributions produces a factorized result. By considering an ensemble of such partition functions differing in the microscopic details, the universal wormhole solution can be made to dominate \cite{saad2021wormholes}.}

\subsubsection*{Wormholes from ensembles}

In low dimensional examples, Euclidean wormholes arise by considering the product of partition functions averaged over an ensemble of theories \cite{Saad:2019lba}. For the more realistic theories that we want to consider, it's not clear that an appropriate ensemble of theories exists. However, as we now explain, it is still possible to have a type of ensemble when considering a single theory. 

As an example, consider a cosmology supported by massive particles moving on geodesic trajectories in the background. In the Euclidean continuation of this cosmology, the massive particle geodesics extend between the asymptotically AdS boundaries at the Euclidean past and the Euclidean future. According to the AdS/CFT dictionary, each geodesic endpoint is associated with the insertion of a corresponding operator in the CFT. Thus, the wormhole should arise from a construction involving many pairs of operator insertions, where the same operator is inserted at corresponding points in the two CFTs. Simply inserting a bunch of operators does not produce any correlations between the CFTs. However, it is natural to consider an ensemble of these insertions where the specific operators being inserted and the locations of insertions vary among members of the ensemble but are correlated between the two CFTs. For the wormhole solution to dominate, the ensemble must contain a sufficiently large amount of correlation between the insertions in the two CFTs. This picture will be discussed in detail in \cite{AMV}.

\subsubsection*{From ensemble to interaction}

\begin{figure}
    \centering
    \includegraphics[scale=0.3]{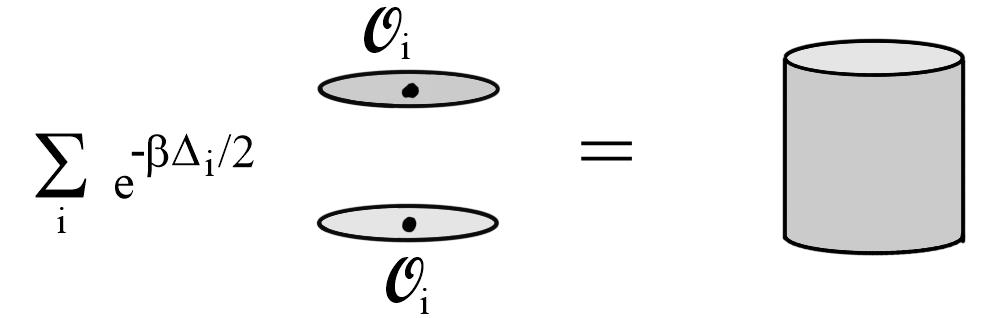}
    \caption{Left: Path integral for a pair of CFTs on ball-shaped regions (depicted as disks) with an ensemble of operator insertions. This constructs the thermofield double state for a pair of CFTs on a sphere. Right: The cylinder path integral also constructs the thermofield double state.}
    \label{fig:TFDtube}
\end{figure}

\begin{figure}
    \centering
    \includegraphics[scale=0.5]{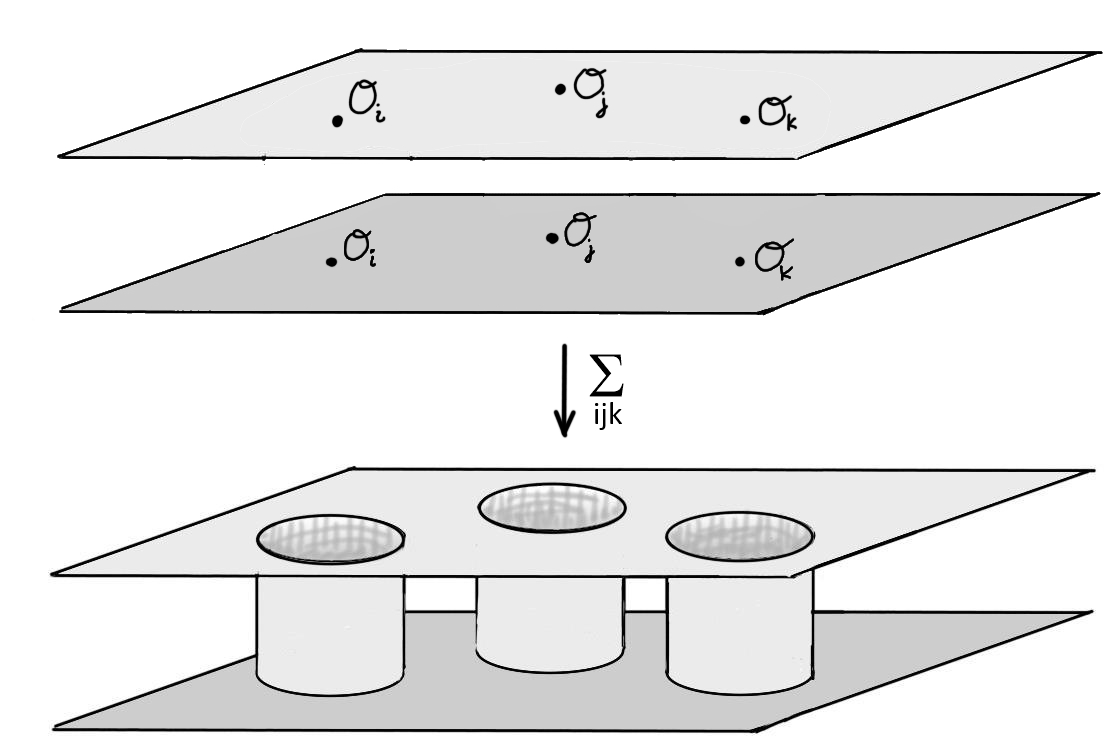}
    \caption{Summing over correlated operator insertions (with a Boltzman factor $\prod_i \exp (-\beta \Delta_i /2)$) is equivalent to connecting the pair of CFTs by cylindrical tubes.}
    \label{fig:Cheese2}
\end{figure}

As a special case of this ensemble of insertions, consider an example where for each pair of insertion points, we have an ensemble with all possible insertions weighted with the Boltzmann factor $\exp (-\beta \Delta_i /2)$. This ensemble of insertions on a pair of disks/balls constructs the thermofield double state of a pair of Lorentzian CFTs on the spatial spheres that are the boundaries of the two balls. The same state is constructed by the path integral on a cylinder, as shown in Figure \ref{fig:TFDtube}. Thus, the ensemble of operator insertions is equivalent to a collection of tubes connecting the two CFTs (Figure \ref{fig:Cheese2}). 

The wormhole saddle is one that topologically fills in the region between the two CFT boundaries but leaves the tubes unfilled. The spatial geometry of the time symmetric slice of this wormhole gives the spatial geometry of the corresponding cosmological solution; this has a spherical boundary for each tube. The interpretation of these boundaries is that each tube 
gives rise to a black hole in the cosmology and each black hole has a second asymptotic region with a spherical asymptotic boundary. So the matter particles in the previous example have been replaced with black holes.

From the CFT point of view, the Euclidean path integral can be understood as preparing an entangled state of a collection of decoupled CFTs living on the disconnected set of spheres that we get by slicing the path integral geometry in half along the time-symmetric slice. The full Lorentzian cosmology is dual to this entangled state. This example will be discussed in detail in \cite{AMV,SV}. For both this construction and the more general ensembles of operator insertions, it is important to understand necessary and/or sufficient conditions for the wormhole saddle to dominate.

In this black hole construction, what started out as an ensemble of operator insertions becomes an explicit interaction between the two CFTs via the geometrical tubes. The final cosmology is encoded in a state of the degrees of freedom associated with the tubes. In the next section, we consider more general ways to introduce an interaction that can introduce correlation between the CFTs and give rise to a wormhole saddle. 

\subsubsection*{Wormholes from interactions}

Starting with the pair of two-dimensional Euclidean CFTs, we can more generally introduce a set of auxiliary degrees of freedom through which the pair of CFTs are coupled \cite{VanRaamsdonk:2020tlr}.

In \cite{Cooper:2018cmb, VanRaamsdonk:2021qgv, Antonini2022}, these auxiliary degrees of freedom were taken to be a four-dimensional quantum field theory living on a slab, with the pair of three-dimensional holographic CFTs coupled to this at the two boundaries of the slab. By taking the auxiliary theory to have relatively few local degrees of freedom, the UV physics of the pair of 3D CFTs is not significantly modified and the dual will still contain two asymptotically AdS regions. But via RG flow, the interaction can produce strong effects in the IR. The idea is that we can induce strong IR correlation between the two CFTs that is reflected geometrically by the two asymptotically AdS boundary regions joining in the interior to give a wormhole.

A particular case of this picture is to take the auxiliary 4D QFT also to be holographic. In this case, the dual spacetime in the construction has a five-dimensional part, with the four dimensional wormhole appearing as an end-of-the-world brane on which gravity is localized. This construction was described originally in \cite{Cooper:2018cmb} and studied further in \cite{Antonini:2019qkt,Waddell:2022fbn,Ross:2022pde} and the very recent \cite{Antonini:2024bbm}.

\subsubsection*{Scalar evolution in the Euclidean models}

\begin{figure}
    \centering
    \includegraphics[scale=0.3]{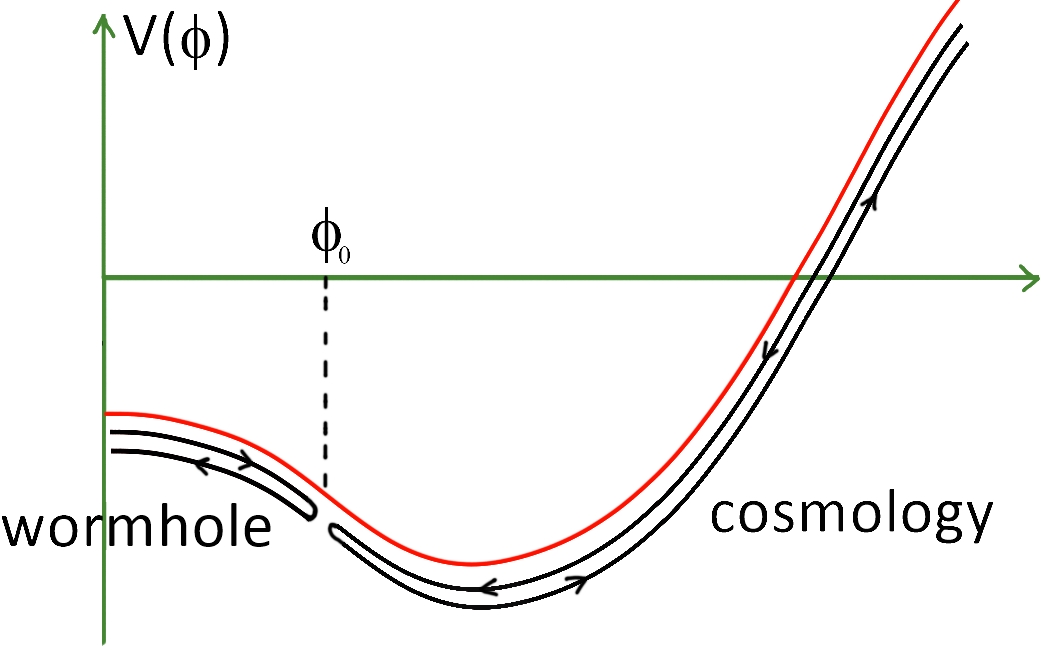}
    \caption{Scalar potential in a field direction corresponding to a relavant operator in the CFT. In the wormhole solution, the scalar evolves from some value $\phi_0$ at the center of the wormhole to $\phi=0$ at the boundaries. In the cosmology. the scalar }
    \label{fig:potential}
\end{figure}

The cosmologies arising from these Euclidean constructions also generically include scalar evolution. The asymptotically AdS regions correspond to an extremum of the scalar potential. As we move away from this boundary, generic solutions will have scalars associated with relevant operators turn on, so we move away from the AdS extremum in a direction along which the scalar potential initially decreases. In time-reflection symmetric wormhole solutions, value of the scalar at the middle of the wormhole also sets the value at the recollapse point of the cosmology, but in the cosmological evolution, the scalar arrives at this point from the other direction. The scalar evolution in the Euclidean wormhole and associated Lorentzian cosmology (which can have an accelerating phase in this example) for a typical potential is shown in Figure \ref{fig:potential}. See \cite{Antonini:2022ptt} for more discussion.

\section{Discussion} \label{sec:dis}

Describing our universe via an effective field theory associated with a holographic CFT is an appealing prospect, since this would likely mean we are far closer to a microscopic model of cosmology than if the observed acceleration is due to a positive cosmological constant.\footnote{See \cite{Betzios:2024oli} for an interesting recent proposal to describe certain $\Lambda > 0$ cosmology models via certain holographic CFTs using a similar approach to the wormhole construction.} Our observational results indicate that time-dependent dark energy, as required in models where a $\Lambda < 0$ EFT produces acceleration, is not only still viable, but may be preferred by current data. It will be very interesting to see how the constraints evolve as additional data become available. To successfully explain observations via a $\Lambda < 0$ EFT with an evolving scalar, many phenomenological questions would need to be addressed. Perhaps most important is the question of why the light scalar apparently doesn't lead to observed long range forces. 

For the microscopic construction, there is significant work to be done to understand how to extract cosmological physics from the CFT in the local picture -- this is related to the general question of understanding the physics behind black hole horizons -- and to understand explicit constructions where we can realize a wormhole saddle in the Euclidean global construction. With either of these microscopic pictures, an important next step is to understand the structure of perturbations and cosmological correlators predicted in the state constructed by the wormhole. We can also try to ask more; whether it is possible and/or natural to realize a period of inflation (or whether the physics of the past singularity itself can produce similar perturbations), whether a classical arrow of time that extends from the big bang to the big crunch can emerge from the time-symmetric quantum state \cite{Antonini2022}, and whether we can learn anything about the nature of the big bang. 

\section*{Acknowledgements}

We thank our collaborators on the work discussed here: Stefano Antonini, Alex Maloney, Viraj Meruliya, Abhisek Sahu, Petar Simidzija, and Brian Swingle. We also thank Douglas Scott and Gary Hinshaw for discussions.  We acknowledge support from the National Science and Engineering Research Council of Canada (NSERC) and the Simons foundation via a Simons Investigator Award. Research at the Perimeter Institute is supported by the Government of Canada
through the Department of Innovation, Science and Industry Canada and by the Province
of Ontario through the Ministry of Colleges and Universities.

\bibliographystyle{JHEP}
\bibliography{refs}

\end{document}